\documentclass[showpacs,pra,preprintnumbers,nofootinbib,aps,floatfix]{revtex4}
\usepackage{epsfig}
\usepackage{subfigure}
\usepackage{graphicx}
\usepackage{natbib}
\usepackage{amssymb}
\usepackage{amsmath}

\newcommand{\six}{\hat{\sigma}_x}
\newcommand{\siz}{\hat{\sigma}_{z}}
\newcommand{\siy}{\hat{\sigma}_y}

\newcommand{\hsig}{\hat{\sigma}}

\newcommand{\hU}{\hat{U}}

\newcommand{\hro}{\hat{\rho}}

\newcommand{\half}{\frac{1}{2}}
\newcommand{\BEQ}{\begin{equation}}
\newcommand{\EEQ}{\end{equation}}
\newcommand{\BEA}{\begin{eqnarray}}
\newcommand{\EEA}{\end{eqnarray}}
\newcommand{\ad}{\hat{a}^{\dagger}}
\newcommand{\add}{\hat{a}}
\newcommand{\spp}{\hat{\sigma}_+}
\newcommand{\smm}{\hat{\sigma}_-}

\newcommand{\nn}{\nonumber \\}
\renewcommand{\thesection}{\arabic{section}}

\begin{document}

\title{Quantum state tomography using a single apparatus}

\author{B. Mehmani$^{1}$, A. E. Allahverdyan$^{2}$, Th. M. Nieuwenhuizen$^{1}$ }
\affiliation{ $^{1}$ Institute for Theoretical Physics,University
of Amsterdam, Valckenierstraat 65, 1018 XE Amsterdam, The
Netherlands} \affiliation{$^{2}$Yerevan Physics Institute,
Alikhanian Brothers St. 2, Yerevan 375036, Armenia}

\begin{abstract} The density matrix of a two-level system (spin, atom)
is usually determined by measuring the three non-commuting components of the
Pauli vector. This density matrix can also be obtained  via
the measurement data of two commuting variables, using a single apparatus. This is done by
coupling the two-level system to a mode of radiation field, where the
atom-field interaction is described  with the Jaynes--Cummings model. The
mode starts its evolution from a known coherent state.  The unknown
initial state of the atom is found by measuring  two commuting observables: the population difference of the
atom and the photon number of the field. We discuss the advantages of
this setup and its possible applications. 

\end{abstract}

\maketitle
\renewcommand{\thesection}{\arabic{section}}
\section{Introduction}
\setcounter{equation}{0}
\renewcommand{\thesection}{\arabic{section}.}

Determining the unknown state of a quantum system is the basic 
inverse problem of quantum mechanics. Given the state, one can
calculate the expectation value of any observable of the system
\cite{deMuynck}. However, the inverse problem of
determining the state by performing different measurements is
non-trivial. This problem was discussed by Pauli \cite{Pauli} in 1933.
His question was whether one can reconstruct the unknown wave-function
of an ensemble of identical spinless particles via the corresponding
position and momentum probability densities. The interest to the state
determination problem grew considerably since then, and now this is a
well-recognized subject \cite{Kemble}. 

The notion of state refers to an ensemble of identically prepared
systems and is represented by a Hermitian operator with non-negative
eigenvalues that sum to one (density matrix) \cite{deMuynck}.
Thus the $N \times N$ density matrix of an ensemble of $N$-level systems is specified
by $N^2 - 1$ independent real parameters. Since any observable
$\hat{\mathcal{O}}$ of the system generates at most $N-1$ independent
probabilities, at least $(N^2 -1)/(N-1) = N+1$ measurements of
non-commuting observables are needed to obtain the unknown state 
\cite{brut}.

Procedures of reconstructing the quantum state from measurements are
known as {\it quantum state tomography}.  Recently they found
applications in quantum information processing \cite{Nielson}. For
example, in quantum cryptography one needs a complete specification of
the qubit state \cite{Pasqui}.  For the simplest example of a
spin-$\half$ system the state is described by a $2 \times 2$ matrix.
According to the above argument, one has to perform $3$ incompatible
measurements for the unknown state determination, e.g., measuring the
spin components along the \textrm{x}-, \textrm{y}- and \textrm{z}- axes
via the Stern-Gerlach setup. However, during the measuremental procedure
of each component one looses the information about the two other
components, since the spin operators in different directions do not
commute. Thus, to determine the state of a spin-$\half$ system, one
needs to use repeatedly three Stern--Gerlach measurements performed
along orthogonal directions. 

However, the state can be characterized indirectly via a single set of
measurements performed simultaneously on the system of interest and an
auxiliary system (assistant), which starts its evolution from a known
state \cite{D'Ariano,PRL,peng}. This can be done by letting them interact
for a specific time and measuring commuting observables of the overall
system (system+assistant). In particular, it has been shown that one
can determine the unknown state of a spin-$\half$ system with a single
apparatus by using another spin-$\half$ assistant \cite{PRL}. The idea was recently implemented by Peng {\it et. al.} \cite{peng} who used pulses to induce the proper dynamics of the interaction between a spin-$\half$ system and its assistant. They verified the initial state of the system obtained from this procedure with the results of the direct measurement of the three components of the spin vector of the system.

Our present purpose is to determine the unknown density matrix
of an ensemble of two-level systems (atom or spin) via interaction with a
single mode of the electromagnetic field. The atom-field interaction is
studied within the Jaynes--Cummings model \cite{Jaynes}. We show that
the unknown state of the spin can be completely characterized by
measuring two commuting variables: the population difference of the
atoms $\hat{\sigma}_z$ and the photon number of the field
$\hat{a}^\dagger \hat{a}$. This measurement supplies three averages:
$\langle\sigma_z\rangle$, $\langle a^\dagger a\rangle$ and
$\langle\sigma_z\, a^\dagger a\rangle$, which will be linearly
related to the elements of the initial density matrix of the ensemble of the two-level atoms. (Note that
since $\hat{\sigma}_z$ and $\hat{a}^\dagger \hat{a}$ commute,
$\langle\sigma_z\,a^\dagger a\rangle$ is recovered from the
measurement data of $\sigma_z$ and $a^\dagger a$ via the number of
coincidences.)

In section \ref{JCM} we will give a brief introduction on the
Jaynes--Cummings model and its properties. In section \ref{procedure}
the model is used to determine the state of the two-level system.  
Section \ref{raduga} discusses imperfection of the proposed scheme due to the
noise in selecting the measurement time. In section \ref{ml} we
discuss how to reconstruct the unknown density matrix approximately
given an incomplete measurement data. The solution of this task
amounts to a direct application of the classical Maximum Likelihood
setup, because our scheme operates with the measurement of commuting
observables. We conclude in section \ref{conclusion}.

\renewcommand{\thesection}{\arabic{section}}
\section{The Jaynes--Cummings Model}\label{JCM}
\setcounter{equation}{0}
\renewcommand{\thesection}{\arabic{section}.}

The Jaynes--Cummings model (JCM) \cite{Jaynes} has a special place in quantum
optics and atomic physics
\cite{Stenholm,Shore,Raimond}. This
model describes the interaction of a two-level atom with a single mode
of electromagnetic field, and it was employed by Jaynes and Cummings for
studying the quantum features of spontaneous emission. Later on, the model generated several
non-trivial theoretical predictions|such as collapes and revivals of the
atomic population that are related to the discreteness of photons
\cite{Cummings,Eberly}|which were successfully tested in experiments
\cite{Haroche}. In particular, the model explains experimental
results on one-atom masers \cite{Haroche}, and on the passage of
(Rydberg) atoms through cavities
\cite{Rempe2,Walther1,Osnaghi,Meunier}. The JCM is also used for
describing quantum correlation and formation of macroscopic quantum
states.  It was recently employed in quantum information theory
\cite{Ellin,JCinfo}. More recently, the JCM found applications in
semiconductors \cite{semiconductor} and in Josephson junctions
\cite{josephson}. 

The JCM is in fact a family of models, since the original model of Jaynes
and Cummings was generalized several times for a more adequate description
of the atom-field interaction (e.g., multi-mode
fields, multi-level atoms, damping) \cite{Kundu,Hussin,Rodri}. We
shall however study the simplest original realization of the JCM that
involves a two-level atom interacting with a single mode of
electromagnetic field. In particular, we neglect the effects of noise
and dissipation. This situation has direct experimental realizations
\cite{Haroche,Garching,Har2}. For instance
with the superconducting microcavities one can achieve $\sim 0.1$s for the
average lifetime of the cavity photon, much larger than the 
typical field-atom interaction time $\sim 100-500\mu s$ \cite{Rempe2, Meunier}.

The quantized mode of the radiation field is described via bosonic creation
$\hat{a}^{\dagger}$ and annihilation $\hat{a}$ operators.  The
two--level system is mathematically identical to a spin-$\half$, and
can be described  with the help of the Pauli matrices $\six$, $\siy$ and $\siz$.  Within
the dipole approximation one has the following Hamiltonian for the atom
and the cavity mode:
\BEA
\hat{H} = \hbar \omega
\hat{\sigma}_z + \hbar \nu \hat{a}^{\dagger} \hat{a} + 
\hbar g (\hat{a}^{\dagger}+\hat{a})\six
\label{koma}
\EEA
where $\omega$ and $\nu$ are the atom frequency and the mode frequency,
respectively, and $g$ is the atom--field coupling constant. In the
dipole approximation it is given as $g = d (\omega/\hbar V
\epsilon_0)^{1/2}$, where $d$ is the atomic dipole matrix element, and
$V$ is the cavity volume. 

We note that in quantum optical realizations of the JCM the coupling constant
$g$ is normally much smaller than $\omega$ and $\nu$, e.g., it is typical to have
$\nu\sim \omega \times 10^{5}$ ($\omega\sim \nu=$10GHz, $g\sim \Delta=$10--100KHz). 
Thus the subsequent reasoning based on the 
interaction representation is legitimate.
To obtain from (\ref{koma}) the JC Hamiltonian note that
in the interaction representation the coupling term reads:
\BEA
\hbar g (\hat{a}^{\dagger}\, e^{-i\nu t}+\hat{a}\, e^{i\nu t})
(\hat{\sigma}_-\, e^{-i\omega t} + \hat{\sigma}_+ e^{i\omega t}),
\label{durban}
\EEA
where we introduced raising 
\BEA
\hat{\sigma}_+ = \six + i \siy,
\EEA
and lowering
\BEA
\hat{\sigma}_- = \six - i \siy,
\EEA
spin operators. Recall that they satisfy
the following commutation rules
\BEQ 
[\hat{\sigma}_{\pm}, \siz] = \mp \hat{\sigma}_{\pm},\,\,\,
[\spp,\smm] = 2 \siz, \,\,\, \spp\smm + \smm \spp = 1.
\EEQ 

One now applies to (\ref{durban}) the rotating wave approximation: the
atom and field frequencies are assumed to be close to each other, and
then the factors $\propto e^{\pm it(\nu+\omega)}$ in (\ref{durban})
oscillate in time stronger than $\propto e^{\pm it(\nu-\omega)}$.  Thus
the factors $\propto e^{\pm it(\nu+\omega)}$ are neglected within this
approximation and one arrives at the JC Hamiltonian:

\BEA
\label{jcm_ham}
\hat{H} = \hbar \omega
\hat{\sigma}_z + \hbar \nu \hat{a}^{\dagger} \hat{a} + \hbar g
(\hat{\sigma}_+ \hat{a} + \hat{\sigma}_- \hat{a}^{\dagger}).
\EEA

We shall denote
\BEA
\Delta = \omega - \nu,
\EEA
for the detuning parameter. For our future purposes we 
consider $\Delta $ as a tunable parameter.
Within the atom-cavity realizations of the JCM, the detuning $\Delta$
can be controlled via the mode frequency $\nu$, that is, by changing the shape of
the cavity. Alternatively, $\Delta$ can be changed via the atom
frequency $\omega$ by applying an electric field across the cavity
\cite{Har83}.  Then $\omega$ is modified due to the Stark effect.

The above standard derivation of (\ref{jcm_ham}) is based on small
detuning $\Delta$ and weak atom-mode coupling $g$:
\BEA
\Delta \ll {\rm min}(\omega, \nu), 
\qquad
g \ll {\rm min}(\omega, \nu).
\EEA
Both these conditions are usually satisfied for quantum optical realizations of the JCM. 

There are however situations|especially for the solid state physics
applications of the Hamiltonian (\ref{koma})|where the atom-field
interaction constant $g$ is not small. To this end it is useful to know
that sometimes the counter-rotating terms $\propto e^{\pm
it(\nu+\omega)}$ vanish due to specific selection rules \cite{crisp},
and then the JCM applies in the strong-coupling situation as well.

The Hamiltonian (\ref{jcm_ham}) is exactly solvable
and the corresponding unitary evolution operator reads \cite{Stenholm}
\begin{eqnarray}
\label{unitary}
&&\hat{U}(t) = e^{- i \nu t (\hat{a}^{\dagger} \hat{a} + \frac{1}{2})}\left( \cos [t \sqrt{\hat{\varphi} + g^2}] - i \Delta/2 \frac{\sin [t  \sqrt{\hat{\varphi} + g^2}]}{\sqrt{\hat{\varphi} + g^2}}\right) |+\rangle\langle+|\nonumber\\
&&- i g e^{- i \nu t (\hat{a}^{\dagger} \hat{a} + \frac{1}{2})} \frac{\sin [t  \sqrt{\hat{\varphi} + g^2}]}{\sqrt{\hat{\varphi} + g^2}} \,\hat{a} |+\rangle\langle-|\nonumber\\
&& -i g e^{- i \nu t (\hat{a}^{\dagger} \hat{a} - \frac{1}{2})} \frac{\sin t \sqrt{\hat{\varphi}}} {\sqrt{\hat{\varphi}}} \,\hat{a}^{\dagger} |-\rangle\langle+|\nonumber\\
&&+ e^{- i \nu t (\hat{a}^{\dagger} \hat{a} - \frac{1}{2})} \left( \cos t \sqrt{\hat{\varphi}} + i \Delta/2 \frac{\sin t \sqrt{\hat{\varphi}}}{\sqrt{\hat{\varphi} }}\right) |-\rangle\langle-|,
\end{eqnarray}
where the operator $\hat{\varphi}$ is defined as
\BEQ
\hat{\varphi} = g^2 \hat{a}^{\dagger} \hat{a} + \Delta^2/4.
\EEQ
The unitarity of $\hat{U}$ is guaranteed by the identities

\begin{eqnarray}\label{unitary condition}
&&\frac{\sin \left[t\,\sqrt{\hat{\varphi} + g^2}\right]}{\sqrt{\hat{\varphi} +
g^2}}\; \hat{a} = \hat{a}\; \frac{\sin\left[
t\,\sqrt{\hat{\varphi} } \right] }{\sqrt{\hat{\varphi}}} ,\nonumber\\
&&\cos \left[t\, \sqrt{\hat{\varphi} + g^2}\right]\; \hat{a} = \hat{a}\; 
\cos \left[t \sqrt{\hat{\varphi}}\right].
\end{eqnarray}

\renewcommand{\thesection}{\arabic{section}}
\section{Determination of the atom initial state.}
\label{procedure}
\setcounter{equation}{0}
\renewcommand{\thesection}{\arabic{section}.}

Let initially the atom be described by some general mixed density matrix
$ \hro_{S}$: 
\BEA
\label{initial}
\hro_{S} = \frac{1}{2}(1+\left\langle \six \right\rangle_0 \six+
\left\langle \siy \right\rangle_0 \siy  + \left\langle \siz \right\rangle_0 \siz ), \qquad 
\left\langle \hat{\sigma}_i \right\rangle_0\equiv {\rm tr}(\hat{\sigma}_i\, \hro_S), \qquad i=x,y,z,
\EEA
where $\left\langle \hat{\sigma}_i \right\rangle_0$ are the three unknown coefficients of the initial atom state.

We shall assume that the mode starts its
evolution from a coherent state with a known parameter $\alpha$:
\BEQ\label{field state}
|\alpha\rangle = e^{- |\alpha|^2/2} \sum_{n=0}^{\infty} \frac{\alpha^n}{\sqrt{n!}}\, |n\rangle,
\EEQ
where $|\alpha\rangle$ is the eigenvector of the annihilation operator $\hat{a}$, 
\BEA
\hat{a}|\alpha\rangle = \alpha|\alpha\rangle,
\EEA
and {where} $|n\rangle$ is the eigenvector of the photon number operator $\hat{a}^{\dagger} \hat{a}$,
\BEA
\hat{a}^{\dagger} \hat{a}|n\rangle = n|n\rangle.
\EEA
For the average number of photons we have:
\BEA
\bar{n}=\langle \alpha|\hat{a}^{\dagger} \hat{a}|\alpha\rangle = |\alpha|^2.
\EEA

The assumption (\ref{field state}) on the initial state of the field is natural since these are
the kinds of fields produced by classical currents \cite{Glauber}, and also, to a
good approximation, by sufficiently intense laser fields.

Since initially the system and the assistant do not interact, the overall initial density matrix is factorized
\BEA
\label{barano}
\label{rho(0)}
\hro(0) = \hat{\rho}_S\otimes |\alpha\rangle \langle \alpha |.
\EEA

With the help of the unitary 
operator (\ref{unitary}) one can
calculate the overall density matrix $\hro(t)$ at time $t$:
\BEA
\label{final}
\hro(t) = \hU(t)\, \hro(0)\, \hU^{\dagger}(t).
\EEA

\begin{figure}[t]
\centering
    \mbox{
       \subfigure[$\bar{n} = 2, \Delta = 10$KHz]{\epsfig{figure=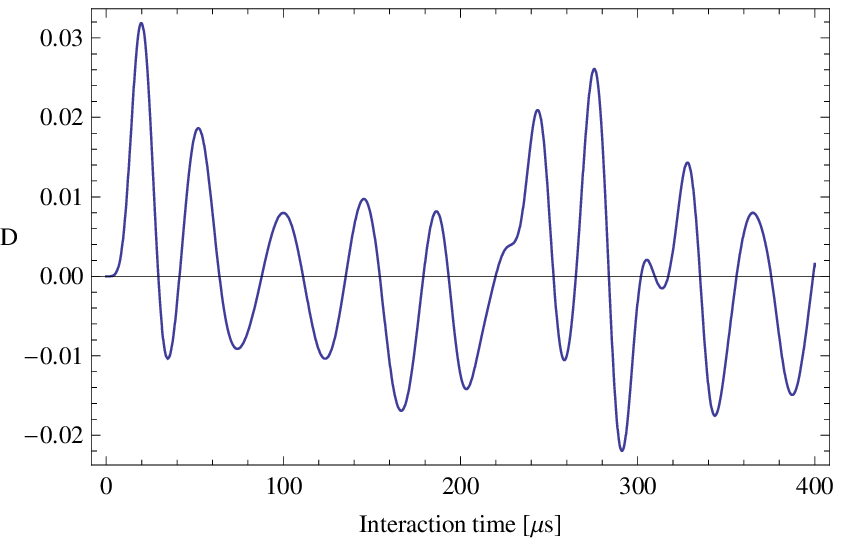,width=2.5in}} \quad
       \subfigure[$\bar{n} = 2, \Delta = 100$KHz]{\epsfig{figure=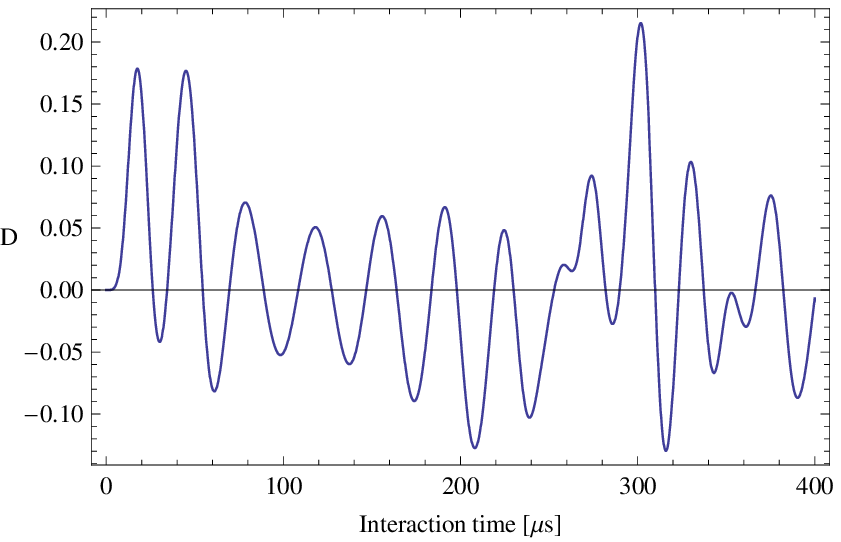,width=2.5in}}
      }
    \mbox{
      \subfigure[$\bar{n} = 5, \Delta = 10$KHz]{\epsfig{figure=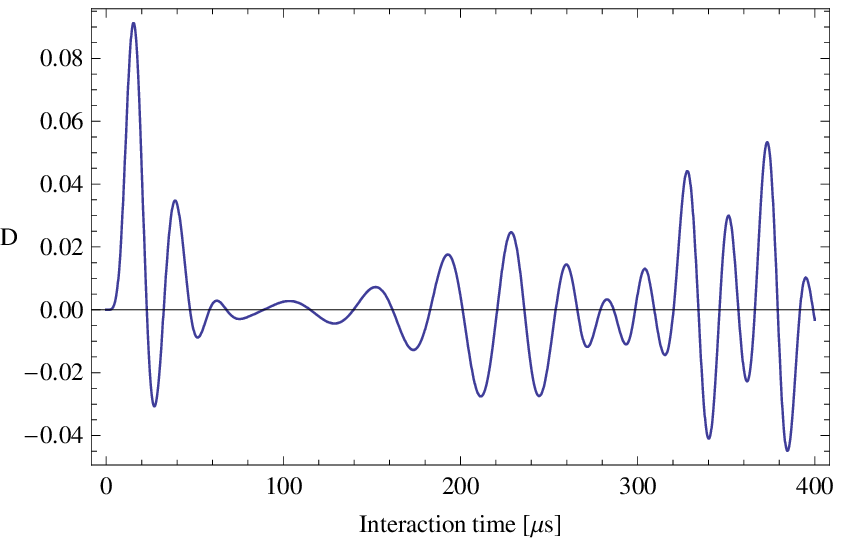,width=2.5in}} \quad
      \subfigure[$\bar{n} = 5, \Delta = 100$KHz]{\epsfig{figure=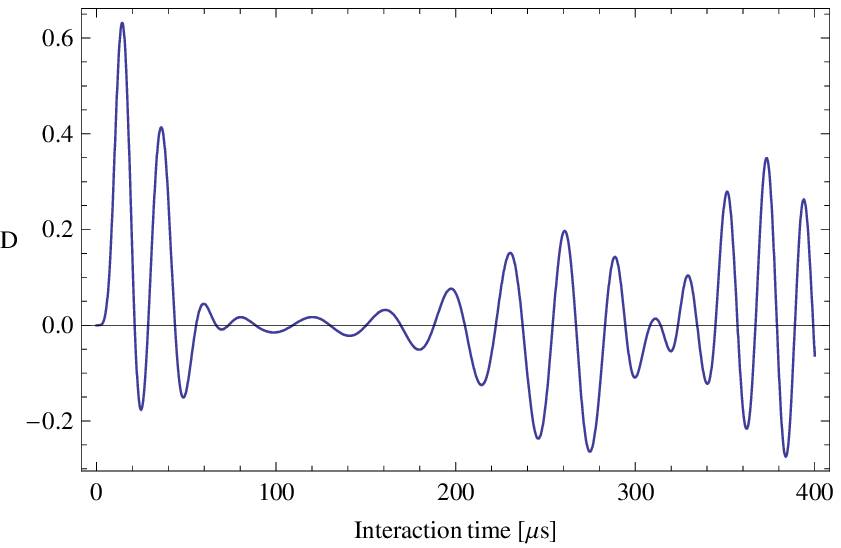,width=2.5in}} 
      }
\mbox{
      \subfigure[$\bar{n} = 10, \Delta = 10$KHz]{\epsfig{figure=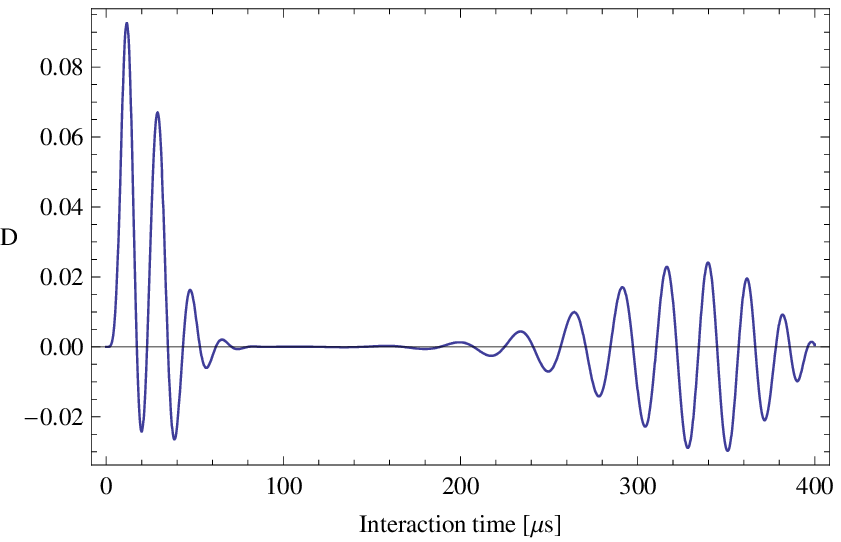,width=2.5in}} \quad
      \subfigure[$\bar{n} = 10, \Delta = 100$KHz]{\epsfig{figure=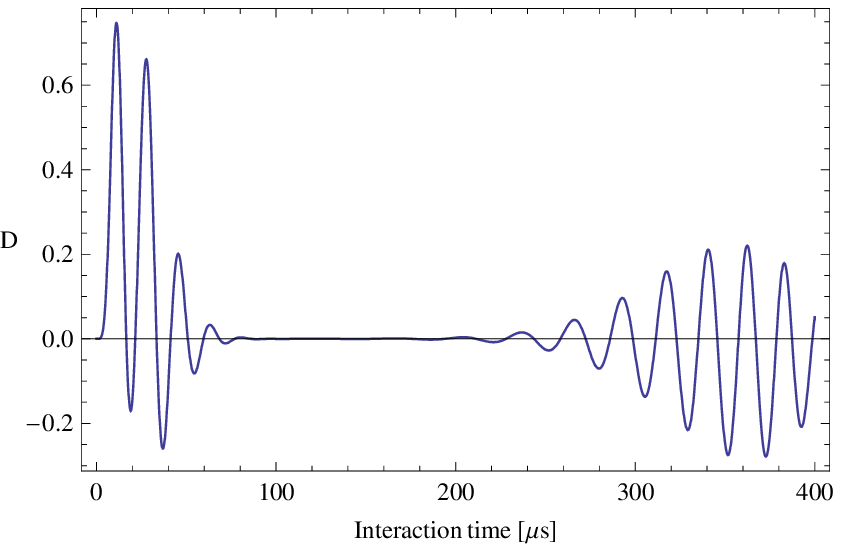,width=2.5in}} 
      }
   \caption{The determinant as a function of time for different values of the mean photon number, $\bar{n}=2$, $\bar{n}=5$, and $\bar{n}=10$ and detuning parameter, $\Delta=10$ and $\Delta=100$ KHz. In all cases $D$ has larger maximum value when $\Delta = 100$KHz. The coupling constant is fixed at $g = 50$KHz.}
\label{detn}
\end{figure}

Then the expectation value of any observable  $\hat{\mathcal{O}}$ of the overall system at time $t$ 
is
\BEQ\label{expectation}
\langle\hat{\mathcal{O}}\rangle = \textnormal{Tr}\left( \hro(t) \hat{\mathcal{O}} \right).
\EEQ

In the Appendix A we work out equations for the atom population difference
$\langle\siz\rangle_t$, the average number of photons
$\langle\ad \add\rangle_t$ and the correlator of these two observables
$\langle\siz \,
\ad \add\rangle_t$; see (\ref{siz}) -- (\ref{siz*n}). 
Expectedly, these three quantities are linearly related to the three unknowns
parameters 
$\langle\six\rangle_0$, $\langle\siy\rangle_0$
and $\langle\siz\rangle_0$ of the initial atom density matrix:
\BEQ\label{matrix}
\begin{pmatrix}
  \langle \siz \rangle_t\\
  \langle \ad \add \rangle_t  \\
  \langle \siz \, \ad \add \rangle_t
\end{pmatrix} = \mathcal{M} \begin{pmatrix}
  \langle \six \rangle_0  \\
 \langle \siy \rangle_0  \\
 \langle \siz \rangle_0
\end{pmatrix}
+\mathcal{B}, \qquad \mathcal{B}\equiv
\begin{pmatrix}
b_1\\
b_2\\
b_2
\end{pmatrix}.
\EEQ
 
The elements of the matrix $\mathcal{M}$ 
and of the vector $\mathcal{B}$
are read off from (\ref{siz}) -- (\ref{siz*n}); see as well Appendix A. They depend
on the parameter $\alpha$ of the initial assistant state, on the parameters $\Delta$ and $g$ of the
JC Hamiltonian, and on the interaction time $t$. Thus, if the determinant of $\mathcal{M}$ is not zero,
one can invert $\mathcal{M}$ and express the unknown parameters of the initial atom density matrix via
known quantities. Although the elements of $\mathcal{M}$ are rather complicated, the
determinant itself is much simpler. It takes the explicit form
\BEA
\label{determinant}
\label{deti}
D\equiv {\rm det}{\mathcal M} =&& 4 \Delta\, g^2 e^{- 2 |\alpha|^2} \sum_{n=0}^{\infty}\sum_{m=0}^{\infty} \frac{|\alpha|^{2(n+m+1)}}{n! m!} (n-m) \times\nn
&&\left[\frac{\sin^2\left( \frac{\Omega_{n}\,t }{2}\right) \,\,\sin  \Omega_{m}\,t}{\Omega_{n}^2 \Omega_{m}} - \frac{\sin^2 \left(\frac{\Omega_{m}\,t}{2}\right)\,\,\sin \Omega_{n}\,t}{\Omega_{m}^2 \Omega_{n} }\right],\nn
\EEA
where $\Omega_n$ is the corresponding Rabi frequency
\BEA
\Omega_n=\sqrt{4(n+1)g^2+\Delta^2}.
\EEA

Eq.~(\ref{determinant}) is our basic result. At the
initial time $t=0$, the determinant $D$ is zero, since the initial state
of the overall system is factorized.  It is seen in
Figs.~\ref{detn}.(a)--\ref{detn}.(f) that for a non-zero detuning,
$\Delta\neq 0$, the determinant $D$ is non-zero for a certain initial
period $t>0$. (Obviously $D$ is zero when there is no photon in the
cavity.) Comparing figures Fig.~\ref{detn}.(a) and Fig.~\ref{detn}.(c)
we see that although higher initial photon numbers $\bar{n}$ lead to
bigger values for the determinant, they cause rapid oscillations in the
value of $D$. This makes the measurement process more difficult. (Note
in this context that the determinant depends on the absolute value of
$\alpha$.)

If the average number of photons $\bar{n}=|\alpha|^2$ in the initial
state of the field is sufficiently large, $D$ collapses to zero for
intermediate times; see Figs.~\ref{detn}.(e) and \ref{detn}.(f). 
The reason for this collapse is apparent from
(\ref{determinant}) and has the same origin as the collapse of the
atomic population difference well known for the JCM \cite{Shore}. Each
term in the RHS of (\ref{determinant}) oscillates with a different
frequency. With time these oscillations get out of phase and $D$
vanishes (collapses).  However, since the number of relevant
oscillations in $D$ is finite, they partially get in phase for later
times producing the revival of $D$, as seen in the
Figs.~\ref{detn}.(e) and \ref{detn}.(f). 

It is seen that $D$ does not depend on separate frequencies $\omega$ and
$\nu$ of the two-level system and the field, only their difference
$\Delta=\omega-\nu$ is relevant. This is due to the choice of the
measurement basis|see the LHS of (\ref{matrix})|that involves quantities
which are constants of motion for $g\to 0$.  Note that $D=0$ for
$\Delta=0$. Thus some non-zero detuning is crucial for the present
scheme of the state determination.  The value of $D$ changes by varying
the detuning parameter $\Delta$.  Comparing the figures
Fig.~\ref{detn}.(a) with Fig.~\ref{detn}.(b), Fig.~\ref{detn}.(c) with
Fig.~\ref{detn}.(d), and Fig.~\ref{detn}.(e) with Fig.~\ref{detn}.(f)
one observes that the value of the highest pick of $D$ increases by an
order of magnitude when the detuning parameter changes from 10KHz to
100KHz.  Note that in Eq.~(\ref{deti}) for the determinant $D$ the
contribution from the diagonal $n=m$ matrix elements of the assistant
initial state $|\alpha\rangle\langle \alpha|$ cancels out. Thus, it is
important to have an initial state of the assistant with non-zero
diagonal elements in the $\{ |n\rangle\}$ basis. 

The basic message of this section is that the determinant $D(t)$ is not
zero for a realistic range of the parameters. This means that the
initial unknown state of the two level system can be determined by
specifying the average atom population difference
$\langle\siz\rangle_t$, the average number of photons $\langle\ad
\add\rangle_t$, and their correlator $\langle\siz \, \ad
\add\rangle_t$. In their turn these quantities are obtained from
measuring two commuting observables: the atom population difference
$\hat{\sigma}_z$ and the photon number $\hat{a}^\dagger \hat{a}$. Having at hand the
proper measurement data for these two observables, one can trivially calculate
$\langle\siz\rangle_t$, $\langle\ad \add\rangle_t$, and find $\langle\siz
\, \ad \add\rangle_t$ via the number of coincidences. 

\subsubsection{Numerical illustration}

At this point it may be instructive to give two concrete 
examples on the inversion of the matrix ${\cal M}$ in (\ref{matrix}).

{\bf 1.} Let us ssume that the average number of photons inside the
cavity is two $\bar{n} =2$, the coupling constant is $g = 50$ KHz, and
the detuning parameter $\Delta = 10$ KHz.  Looking at Fig.~1(a) one sees
that the determinant is maximal at (approximately) $\tau =20 \mu s$.
(Recall that the typical interaction time of a thermal atomic beam with
the single mode of the field is of the order of $100\mu s$ \cite{Rempe2,
Meunier}.) The elements of the matrtix ${\cal M}$ and the vector ${\cal
B}$ are worked out in Appendix.  Inserting all these numbers into
(\ref{siz}) - (\ref{a3}) one obtains
\BEA\label{gis1}
{\cal M}^{-1}|_{\Delta=10KHz} = \left(
\begin{matrix}
 15.183 & 5.59578 & 0.0456968 \\
 1.14077 & -1.3668 & -1.38923 \\
 1 & 1 & 0
                \end{matrix}
\right),
\EEA
\BEA\label{gis2}
{\cal B}|_{\Delta=10KHz} = \left(
\begin{matrix}
- 0.0557631\\
2.05576\\
0.0411884
\end{matrix}
\right).
\EEA

{\bf 2.} For the second example we take a larger detuning: $\bar{n} =2$,
$g = 50$ KHz and $\Delta = 100$ KHz.  The proper interaction time $t=
300 \mu s$ is read off from Fig.~1(b) (interaction time
$t\approx 18\,{\rm \mu s}$ gives somewhat smaller determinant; see Fig.~1(b)). The numerical calculation of
${\cal M}^{-1}$ and ${\cal B}$ produces:
\BEA\label{gis3}
{\cal M}^{-1}|_{\Delta=100KHz} = \left(
\begin{matrix}
 3.18085 & 1.23251 & -1.15186 \\
 5.92052 & -4.20194 & -4.52702 \\
 1 & 1 & 0
                \end{matrix}
\right),
\EEA
\BEA\label{gis4}
{\cal B}|_{\Delta=100KHz} = \left(
\begin{matrix}
- 0.0707962\\
2.0708\\
-0.119635
\end{matrix}
\right).
\EEA

\subsection{Random interaction time.}
\label{raduga}

We saw above that the success of the presented scheme is to a large extent
determined by the ability to select properly the interaction time $t$,
since this ultimately should ensure a non-zero (and sufficiently
large) determinant $D$ (It is clear that a small determinant will amplify
numerical errors; see the next section for an example). 

To quantify the robustness of the presented scheme 
it is reasonable to assume that
there is no perfect control in choosing the interaction time.
To this end let us assume
that the interaction time $t$ is a random, Gaussian distributed
quantity centered at $t_0$ with a
dispersion $\sigma$. The corresponding probability distribution $P(t)$ of
thus reads
\BEA
P(t)=\frac{1}{2\pi\sigma}\, e^{-(t-t_0)^2/(2\sigma)}.
\label{tf}
\EEA
One now averages the determinant $D$ over this distribution,
\BEA
\label{gomesh}
\overline{D}(t_0)
 =&& 4 \Delta\, g^2 e^{- 2 |\alpha|^2}
\sum_{n=0}^{\infty}\sum_{m=0}^{\infty} \frac{|\alpha|^{2(n+m+1)}}{n!
m!} (n-m) 
\left[
w(\Omega_n,\Omega_m;t_0)-w(\Omega_m,\Omega_n;t_0)
\right],
\label{kaku1}
\EEA
where
\BEA
\label{kaku2}
&& w(\Omega_n,\Omega_m;t_0)=\nonumber\\
&&\frac{1}{4\Omega_n^2\Omega_m}\,\left[2 e^{-\frac{\sigma}{2}{\Omega_m}^2 }\, \sin[t_0 \Omega_m ] - e^{-\frac{\sigma}{2}(\Omega_m+\Omega_n)^2} \sin[t_0 (\Omega_m+\Omega_n)] - 
e^{-\frac{\sigma}{2}(\Omega_m-\Omega_n)^2} \sin[(\Omega_m-\Omega_n) t_0]\right]
\EEA

It is seen that the oscillations of $D(t)$ turn after averaging to
exponential factors $e^{-\frac{\sigma^2}{2}(\Omega_m \pm \Omega_n)^2}$
and $e^{-\frac{\sigma^2}{2}\Omega_m^2}$ in (\ref{kaku1}, \ref{kaku2}),
due to which the averaged determinant $\overline{D}(t_0)$ gets
suppressed for a sufficiently large ``indeterminacy'' $\sigma$. This suppression is
illustrated in Figs.~2 (a) and 2 (b).  

\begin{figure}[t]
\centering
    \mbox{
       \subfigure[$\bar{n} = 2, \Delta = 100$KHz, $\sigma = 0.01 \mu s$]{\epsfig{figure=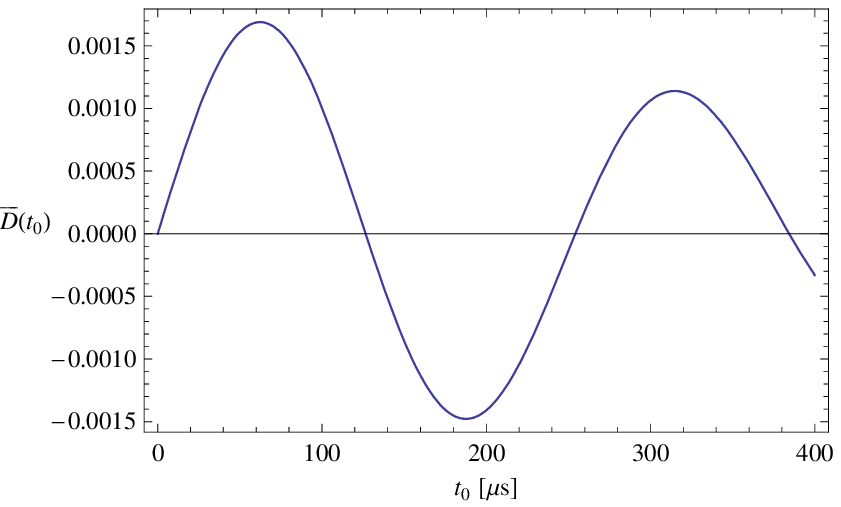,width=2.5in}} 
      }
    \mbox{
            \subfigure[$\bar{n} = 2, \Delta = 10$KHz, $\sigma = 0.1 \mu s$]{\epsfig{figure=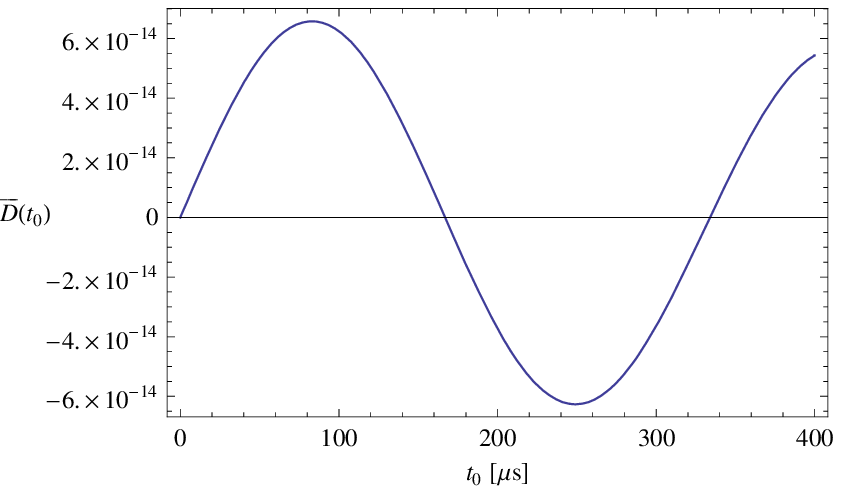,width=2.5in}} 
      }
\caption{The averaged determinant $\bar{D}$
as a function of $t_0$ for different values of $\sigma$; see (\ref{tf}--\ref{kaku2}).
Here $g = 50$KHz.}
\label{detave}
\end{figure}

\renewcommand{\thesection}{\arabic{section}}
\section{Maximum likelihood reconstruction of the initial spin state.}
\setcounter{equation}{0}
\renewcommand{\thesection}{\arabic{section}.}\label{ml}

Above we have shown how one determines the initial spin density matrix
given the three averages $\langle\siz\rangle_t$, $\langle\ad
\add\rangle_t$, and $\langle\siz \, \ad \add\rangle_t$.  However, in
practice the measurement statistics that leads to the above three
averages may be incomplete (due to various noises and experimental
imperfections) and we should understand how to reconstruct the initial
density matrix approximately given the incomplete measurement data. The
general question on the approximate state reconstruction (given incomplete statistics) is of obvious
importance and it got much attention in the standard schemes of quantum
tomography; see \cite{hradil} and references therein. Recall that in these
schemes one measures non-commuting observables. In this context
Refs.~\cite{hradil} propose to generalize suitably the classical Maximum
Likelyhood (ML) method; see \cite{rao} for a detailed discussion on this
inference method. In particular, this generalization accounts for the
fact that the (incomplete) data is obtained from measuring non-commuting
observables. 

Once our scheme operates by measuring the commuting variables only, we
are going to show that for the approximate state reconstruction one now
does not need anything beyond the most standard (classical) ML
method. Since one measures the number of photons and the
spin direction along the $z$-axes (these quantities are represented by
the operators $\ad \add$ and $\siz$, respectively), the incomplete
data in our case means that we are given frequencies $\nu_\alpha
(m)$ of events, where one registered $m$ photons ($m=0,1,\ldots$), and
where, simultaneously, the spin component assumed values $\alpha=\pm
1$. Now recall \cite{rao} that in the ML method the probabilities 
$p_\alpha(m)$ (given the frequencies $\nu_\alpha(m)$)
are obtained by maximizing over $p_\alpha(m)$ the likelihood function \footnote{Equivalently, one
can mimize over $p_\alpha(m)$ the relative entropy $\sum_{\alpha=\pm 1}\sum_{m=0}^\infty
\nu_\alpha(m)\ln \frac{\nu_\alpha(m)}{p_\alpha(m) }$. This measure of distinguishability between 
$p_\alpha(m)$ and $\nu_\alpha(m)$ is equal to zero if and only if $p_\alpha(m)=\nu_\alpha(m)$ and it has
an important information-theoretic meaning \cite{rao}. }
\BEA
\label{mumu}
{\mathcal L}[p_\alpha(m)]=\sum_{\alpha=\pm 1}\sum_{m=0}^\infty
\nu_\alpha(m)\ln p_\alpha(m).
\EEA
This maximization over $p_\alpha(m)$ is to
be carried out in the presence of relevant constraints. For our case 
the initial spin density
matrix $\rho_S$ must be a positive-definite, normalized matrix; see (\ref{barano}).
Thus we get a single constraint
\BEQ
  \langle \six \rangle_0^2+
 \langle \siy \rangle_0^2+
 \langle \siz \rangle_0^2\leq 1.
\EEQ
Working out (\ref{matrix}) we write this constraint as a function of
the probabilities $p_\alpha(m)$:
\BEA
\label{bala}
(u- \mathcal{B})^T \mathcal{C} (u- \mathcal{B}) \leq 1,
\EEA
where $^T$ means transposition, 
${\mathcal C}\equiv ({\mathcal M}{\mathcal M}^T)^{-1}$, the matrix ${\mathcal M}$ and
the vector $\mathcal{B}$ are defined in (\ref{matrix}), and where finally
\BEA
\label{gady}
u=\begin{pmatrix} 
\sum_{\alpha=\pm 1}\sum_{m=0}^\infty \alpha p_\alpha(m)\\ \\
\sum_{\alpha=\pm 1}\sum_{m=0}^\infty m p_\alpha(m)\\ \\
\sum_{\alpha=\pm 1}\sum_{m=0}^\infty \alpha m p_\alpha(m)
\end{pmatrix} .
\EEA
If the constraint (\ref{bala}) is satisfied automatically, the
maximization of ${\mathcal L}[p_\alpha(m)]$ in (\ref{mumu}) produces
\cite{rao}
\BEA
\label{kaban}
p_\alpha(m)=\nu_\alpha(m),
\EEA
i.e., that the sought probabilities are equal to the frequencies, as one would expect
intuitively.
However, in general this constraint is not satisfied automatically and
has to be included explicitly in the maximization of ${\mathcal
L}[p_\alpha(m)]$ over $p_\alpha(m)$. Indeed, looking at (\ref{mumu}) and
(\ref{bala}) we may deduce qualitatively that the constraint
(\ref{bala}) will be satisfied automatically by (\ref{kaban}), if the
frequencies are not very far from the actual probabilities (the ones
that would be obtained in the perfect experiment) and, simultaneously,
the determinant ${\rm det}{\mathcal M}$ is not very close to zero. 
 
\subsubsection{Numerical illustration}

Below we give a concrete numerical example, where the constraint
(\ref{bala}) may or may not be satisfied automatically. We take
$\bar{n}=2$, $g=50$ kHz, $\Delta=100$ kHz, and we have chosen the
measurement time $t= 300 {\rm \mu s}$ such that the corresponding
determinant $D$ is maximized; see Fig. 1(b). Then we constructed the
matrix $\mathcal{C}$ and the vector $\mathcal{B}$ in (\ref{bala}) [see
(\ref{gis3}, \ref{gis4})], and neglecting the probabilities of having
more than three photons inside the cavity, we assumed that we are given
the following six frequencies $\nu_{\pm 1}(m)$ ($m=1,2,3$) normalized
according to $\sum_{m=1}^3 \sum_{\alpha=\pm 1}\nu_\alpha(m)=1$. For
simplicity we additionally assume that these frequencies are related as
\BEA
\nu_{1}(1)=\nu_{-1}(1), \quad
\nu_{1}(2)=\nu_{-1}(2), \quad
\nu_{1}(3)=\nu_{-1}(3).
\label{rabat}
\EEA
For different values of $\nu_{1}(1)$, $\nu_{1}(2)$ and $\nu_{1}(3)$ the numerical
maximization of (\ref{mumu}) over $p_\alpha (m)$ under the constraint (\ref{bala})
produced a result different from (\ref{kaban}) ($\sum_{m=1}^3 \sum_{\alpha=\pm 1}p_\alpha(m)=1$). 
An example follows: for 
\BEA
\label{gr}
\nu_{1}(1) = 0.05, \quad
\nu_{1}(2) = 0.25, \quad 
\nu_{1}(3) = 0.2 
\EEA
the probabilities are:
\BEA
 &&p_{1}(1)=0.05148118,\quad p_{-1}(1)=0.05087771,\nonumber\\
 &&p_{1}(2)=0.24811809,\quad p_{-1}(2)=0.254403426, \nonumber\\
 &&p_{1}(3)=0.19158279,\quad p_{-1}(3)=0.20353679.
\label{mexico}
 \EEA

Employing these probabilities in (\ref{gady}) and in (\ref{matrix}) we get for the initial spin density matrix:
\begin{equation}\label{positive rho}
\hat{\rho}_S = \frac{1}{2} \left[ 1 - (0.187183)\, \hat{\sigma}_x - (0.942992)\, \hat{\sigma}_y + 
(0.275121)\, \hat{\sigma}_z \right]. 
\end{equation}

In this context we need to quantify the difference between the input
frequencies $\nu_\alpha(m)$ and the probabilities $p_\alpha(m)$ which
result from maximizing (\ref{mumu}) under the constraint (\ref{bala}).
In particular, this difference will quantify the relevance of the
constraint (\ref{bala}) in maximizing (\ref{mumu}).  A good measure of distance between two
probability sets is provided by \cite{rao}
\begin{equation}
 \delta [\nu||p] = 1 - \sum_{\alpha = \pm 1} \sum_{m = 1}^{3} \sqrt{\nu_{\alpha}(m) p_{\alpha}(m)}.
\label{ddd}
\end{equation}
This quantity is equal to its minimal value zero if (and only if)
$\nu_\alpha(m)=p_\alpha(m)$ (i.e., when the constraint (\ref{bala}) holds automatically), and it is
equal to its maximal value $1$ for $\nu_\alpha(m)p_\alpha(m)=0$.

In Table~I we calculated the distance $ \delta [\nu||p]$ between the
frequencies and the corresponding probabilities.  It is seen that in
some cases this distance is just equal to zero, while for other cases
it is rather small. 
 
\begin{table}[t]
\centering
\begin{tabular}{|c|c|c|c|c|c|c|}  
\hline
&$\nu_{1}(1)$ = 0.05 & $\nu_{1}(1)$ = 0.15 & $\nu_{1}(1)$ = 0.25 & $\nu_{1}(1)$ = 0.30\\
\hline
$\nu_{1}(2)$ = 0.05 & $\delta$ = 0.00989504 & $\delta$ = 0.0000494347 & $\delta$ = 1.1102 $\times 10^{-16}$  & $\delta$ = 0.00108428\\  
\hline
$\nu_{1}(2)$ = 0.15 & $\delta$ = 0.00318619  & $\delta$ = 0 & $\delta$ = 0.00140516 & $\delta$ = 0.0115233\\
\hline
$\nu_{1}(2)$ = 0.25 & $\delta$ = 0.0000717018 & $\delta$ = 0 & $\delta$ = 0.0336704 &  \\
\hline
$\nu_{1}(2)$ = 0.30 & $\delta$ = 1.1102 $\times 10^{-16}$ & $\delta$ = 0.0000961022 &   &  \\
\hline
\end{tabular}
\caption{The distance $\delta [\nu||p]$ (given by (\ref{ddd}))
between the set of frequencies $\nu_{\pm 1}(m)$ and the set of corresponding probabilities $
p_{\pm 1} (m)$ obtained from maximizing (\ref{mumu}) under the constraint (\ref{bala}). 
As in the main text, we assumed that the frequencies satisfy (\ref{rabat}). 
Thus the third frequency (not shown in the table) is got from $\mu_\alpha(3)=\frac{1}{2}
-\mu_\alpha(1)-\mu_\alpha(2)$. The numerical
values for the involved parameters are: $\bar{n}=2$ (the initial average number of photons), $g=50$ kHz
(coupling constant), $\Delta=100$ kHz (detuning) and $t = 300 {\rm \mu s}$ (the interaction time).
The matrix $\mathcal{M}$ and the vector $\mathcal{B}$ in this case are given by (\ref{gis3}, \ref{gis4}).
Three places in the table are empty, because the corresponding frequencies are unphysical (their sum is
larger than one due to the assumption (\ref{rabat})).
}
\label{math500grades}
\end{table}

\renewcommand{\thesection}{\arabic{section}}
\section{Conclusion}
\setcounter{equation}{0}
\renewcommand{\thesection}{\arabic{section}.}\label{conclusion}

In this paper we describe a method for quantum state tomography.
The usual way of solving this inverse problem of quantum mechanics is to
make measurements of non-commuting quantities. Single apparatus
tomography proceeds differently employing controlled interaction and
measuring commuting observables. This is done via coupling the
system of interest to an auxiliary system (assistant) that starts its
evolution from a known state. The essence of the method is that the proper
coupling is able to transfer the information on the initial state of the system 
to a commuting basis of observables for the composite system (system+assistant). 

It is important to implement the single-apparatus tomography for a
situation with a physically transparent measurement base and with a
realistic system-assistant interaction. Here we carried out this program
for a two-level atom (system) interacting with a single mode of
electromagnetic field (assistant). The atom-field interaction is given
by the Jaynes-Cummings Hamiltonian, which has direct experimental
realizations in quantum optics
\cite{Raimond,Haroche,Rempe2,Walther1,Osnaghi,Meunier},
superconductivity \cite{josephson}, semiconductor physics
\cite{semiconductor}, etc. As the measurement base we have taken
presumably the simplest set of observables related to the energies of
the atom and field: population difference of the atoms $\hat{\sigma}_z$ and the
number of photons $\hat{a}^\dagger \hat{a}$ in the field.  We have shown that one
can determine the unknown initial state of the atom via post-interaction
values of the average atomic population difference $\langle
\hat{\sigma}_z\rangle$, the average number of photons $\langle \hat{a}^\dagger
\hat{a}\rangle$ and the correlator of these quantities $\langle\hat{\sigma}_z\,
\hat{a}^\dagger \hat{a}\rangle$. The latter quantity does not need a separate measurement,
since it can be recovered from the simultaneous measurement of the two
basic observables $\hat{\sigma}_z$ and $\hat{a}^\dagger \hat{a}$. 

Since our scheme is based on measuring commuting observables, we can
apply (more or less literally) the standard Maximum Likelihood setup
for an approximate reconstruction of the unknown density matrix given
the incomplete (noisy) measurement data. This is discussed in section 
\ref{ml}.

\section*{Acknowledgment}
The authors are grateful for discussion with Robert Spreeuw.
A. E. Allahverdyan is supported by Volkswagenstiftung
(grant ``Quantum Thermodynamics: Energy and information flow at nanoscale'') and acknowledges hospitality at the University of Amsterdam. His work was partially supported by the Stichting
voor Fundamenteel Onderzoek der Materie (FOM, financially
supported by the Nederlandse Organisatie voor Wetenschappelijk Onderzoek,
NWO).

\appendix

\section{Derivation of Eq.~(\ref{deti}).}
\label{appendix A}

Here we shall derive formulas for $\langle\siz\rangle_t$, $\langle\ad
\add\rangle_t$ and $\langle\siz \, \ad \add\rangle_t$ starting from
(\ref{initial}, \ref{final}, \ref{unitary}). They are used in deriving
(\ref{deti}) and they define the matrix $\mathcal{M}$ in (\ref{matrix}). 
One derives after straightforward algebraic steps

\BEA\label{siz}
&&\langle\siz\rangle_t = \nn
&& - 2 i g \langle\six\rangle_0 \sum_{n=0}^{\infty}  c_n \frac{ \sin \left( \frac{\Omega_{n}\,t }{2}\right)} {\Omega_{n}}\left \{\alpha\,\left[\cos \left( \frac{\Omega_{n}\,t }{2}\right) + i \Delta\,\frac{\sin \left( \frac{\Omega_{n}\,t }{2}\right)} {\Omega_{n}}\right]  - c.c\right\}\nn
&&+ 2 g \langle\siy\rangle_0 \sum_{n=0}^{\infty}  c_n \frac{\sin \left( \frac{\Omega_{n}\,t }{2}\right)} {\Omega_{n}}\left\{\alpha\,\left[\cos \left( \frac{\Omega_{n}\,t }{2}\right) + i \Delta\,\frac{\sin \left( \frac{\Omega_{n}\,t }{2}\right)} {\Omega_{n}} \right] + c.c\right \}\nn
&&+ \langle\siz\rangle_0 \left\{ 1 - g^2\, \sum_{n=0}^{\infty} (n+1)( c_{n+1} + c_n )\, \frac{\sin^2\left( \frac{\Omega_n\,t}{2} \right)}{\left( \frac{\Omega_n}{2} \right)^2} \right\}
+b_1, \nn
\EEA
where $\langle\hsig_i\rangle_0$, $i = x, y, z$, 
are the unknown elements of the initial atom density matrix, $(X+{\it c.c.})$ stands for 
$(X+X^*)$,  and where $b_1$, $c_n$ and $\Omega_n$ are defined as
\BEA\label{b1}
&&b_1\equiv\frac{g^2}{2} \sum_{n=0}^{\infty} (n+1) (c_{n+1} - c_n)\,\frac{\sin^2\left( \frac{\Omega_n\,t}{2} \right)}{\left( \frac{\Omega_n}{2} \right)^2}
, \nn
\EEA
\BEQ
c_n \equiv e^{- |\alpha|^2}\,\frac{\alpha^{2n}}{n!},
\EEQ
\BEQ
{\Omega_n}^2 \equiv 4 (n + 1) g^2 + \Delta^2.
\EEQ

The average number of photons in the cavity, $\ad \add$, can be calculated in a similar way
\BEA\label{n}
&&\langle\ad \add\rangle_t =\nn
&& 2 i g \langle\six\rangle_0 \sum_{n=0}^{\infty}  c_n \frac{\sin \left( \frac{\Omega_{n}\,t }{2}\right)} {\Omega_{n}}\left\{\alpha\,\left[\cos \left( \frac{\Omega_{n}\,t }{2}\right) + i \Delta\,\frac{\sin \left( \frac{\Omega_{n}\,t }{2}\right)} {\Omega_{n}} \right] - c.c\right\}\nn
&&- 2 g \langle\siy\rangle_0 \sum_{n=0}^{\infty}  c_n \frac{\sin \left( \frac{\Omega_{n}\,t }{2}\right)} {\Omega_{n}}\left\{\alpha\,\left[\cos \left( \frac{\Omega_{n}\,t }{2}\right) + i \Delta\,\frac{\sin \left( \frac{\Omega_{n}\,t }{2}\right)} {\Omega_{n}}\right]  + c.c\right\}\nn
&& + g^2 \langle\siz\rangle_0 \sum_{n=0}^{\infty}  (n+1) (c_{n+1} + c_n)\,\frac{\sin^2\left( \frac{\Omega_n\,t}{2} \right)}{\left( \frac{\Omega_n}{2} \right)^2}+ b_2,\nn
\EEA
where $b_2$ is defined as
\BEA\label{b2}
&&b_2\equiv \sum_{n=0}^{\infty} n c_n - \frac{g^2}{2}  \sum_{n=0}^{\infty} (n+1) (c_{n+1} - c_n) \,\frac{\sin^2\left( \frac{\Omega_n\,t}{2} \right)}{\left( \frac{\Omega_n}{2} \right)^2}.\nn
\EEA

The correlator of the two observables reads
\BEA\label{siz*n}
&&\langle\siz \otimes \ad \add\rangle_t =\nn
&&- i g \langle\six\rangle_0
\sum_{n=0}^{\infty}  c_n(2 n + 1)  \frac{\sin \left( \frac{\Omega_{n}\,t }{2}\right) }{\Omega_{n}} \left\{\alpha\left[\cos \left( \frac{\Omega_{n}\,t }{2}\right) + i \Delta \frac{\sin \left( \frac{\Omega_{n}\,t }{2}\right) }{\Omega_{n}}\right] - c.c\right\}\nn
&&+ g \langle\siy\rangle_0 \sum_{n=0}^{\infty} c_n(2 n + 1)  \frac{\sin \left( \frac{\Omega_{n}\,t }{2}\right) }{\Omega_{n}} \left\{ \alpha\left[\cos \left( \frac{\Omega_{n}\,t }{2}\right) + i \Delta \frac{\sin \left( \frac{\Omega_{n}\,t }{2}\right) }{\Omega_{n}} \right] + c.c\right\}\nn
&&+ \langle\siz\rangle_0 \left\{ \sum_{n=0}^{\infty} n c_n - \frac{(n+1) g^2}{2} \left[ (2n+3)c_{n+1} + (2 n +1) c_n\right]\,\frac{\sin^2\left( \frac{\Omega_n\,t}{2} \right)}{\left( \frac{\Omega_n}{2} \right)^2} \right\} + b_3,\nn
\EEA
where $b_3$ is defined as
\BEA\label{a3}
&&b_3\equiv \frac{g^2}{4} \sum_{n=0}^{\infty}  (n+1) \left[ (2n +3) c_{n+1} - (2n+1)c_n\right]\,\frac{\sin^2\left( \frac{\Omega_n\,t}{2} \right)}{\left( \frac{\Omega_n}{2} \right)^2} .\nn
\EEA

\end{document}